\documentclass{aa}
\usepackage{psfig}

\def\la{\;
\raise0.3ex\hbox{$<$\kern-0.75em\raise-1.1ex\hbox{$\sim$}}\; }
\def\ga{\;
\raise0.3ex\hbox{$>$\kern-0.75em\raise-1.1ex\hbox{$\sim$}}\; }

\newcommand{\zabs}{$z_{\rm abs}\,$}
\newcommand{\zem}{$z_{\rm em}\,$}

\newcommand{\kms}{km~s$^{-1}\,$}
\newcommand{\ms}{m~s$^{-1}\,$}
\newcommand{\cm}{cm$^{-2}\,$}

\newcommand{\daa}{$\Delta\alpha/\alpha\,$}

\begin{document}

\title{VLT/UVES constraints on the cosmological variability of 
the fine-structure constant\thanks{Based on observations performed 
at the VLT Kueyen telescope (ESO, Paranal, Chile). The data are retrieved  
from the ESO/ST-ECF Science Archive Facility.
}
}

\author{
S. A. Levshakov\inst{1}
\and
M. Centuri\'on\inst{2}
\and
P. Molaro\inst{2}
\and
S. D'Odorico\inst{3}
}

\offprints{S.~A.~Levshakov
\protect \\lev@astro.ioffe.rssi.ru}

\institute{
Department of Theoretical Astrophysics,
Ioffe Physico-Technical Institute, 194021 St.Petersburg, Russia
\and
Osservatorio Astronomico di Trieste, Via G. B. Tiepolo 11,
34131 Trieste, Italy
\and
European Southern Observatory, Karl-Schwarzschild-Strasse 2,
D-85748 Garching bei M\"unchen, Germany
}

\date{Received 00  / Accepted 00 }

\abstract{We propose a new methodology for probing the cosmological
variability of $\alpha$ from pairs of \ion{Fe}{ii} lines 
(SIDAM, single ion differential $\alpha$ measurement)  
observed in individual exposures from a high resolution spectrograph. 
By this we avoid the influence of the spectral shifts due to
$(i)$ ionization inhomogeneities in the absorbers and 
$(ii)$ non-zero offsets between different exposures. 
Applied to the \ion{Fe}{ii} lines of the metal absorption line system at 
\zabs = 1.839 in the spectrum of 
\object{Q 1101--264} obtained by means of 
the UV-Visual Echelle Spectrograph (UVES) at the ESO Very Large Telescope (VLT),
SIDAM provides
\daa = $(2.4\pm3.8_{\rm stat})\times10^{-6}$.
The \zabs = 1.15 \ion{Fe}{ii} system toward \object{HE 0515--4414}
has been re-analyzed by this method
thus obtaining for the combined sample 
\daa = $(0.4\pm1.5_{\rm stat})\times10^{-6}$.
These values are shifted with respect to  the Keck/HIRES mean
\daa = $(-5.7\pm1.1_{\rm stat})\times10^{-6}$ 
(Murphy et al. 2004) at very high confidence level (95\%). 
The fundamental photon noise limitation in the 
\daa measurement with the VLT/UVES is discussed 
to figure the prospects for future observations. 
It is suggested that with a spectrograph of 
$\sim$\,10 times the UVES dispersion coupled to a 100~m class telescope
the present Oklo level (\daa $\geq 4.5\times10^{-8}$) can be achieved 
along cosmological distances with differential measurements of \daa.
} 
\authorrunning{S. A. Levshakov et al.}
\titlerunning{VLT/UVES constraints on the variability of $\alpha$}
\maketitle

\section{Introduction}

The variability of the fundamental physical constants  over cosmic time, 
firstly proposed almost seven decades ago (Milne 1937; Dirac 1937), has been
subsequently investigated in various aspects by many authors
(for a review see, e.g., Uzan 2003).

The Sommerfeld fine-structure constant, 
$\alpha \equiv e^2/\hbar c$, which describes 
electromagnetic and optical properties of atoms, is the most suitable 
for time variation tests in both laboratory experiments with atomic clocks and 
astronomical observations. The value of this constant is known with
high accuracy, 
$\alpha = 1/137.035\,999\,76(50)$ (Mohr \& Tailor 2000), and its
time-dependence is restricted in the laboratory experiments at
the level of $d$ln[$\alpha(t)]/dt = (-0.9\pm2.9_{\rm stat})\times10^{-15}$ yr$^{-1}$,
corresponding to an upper limit of
$|d{\rm ln}[\alpha(t)]/dt| < 3.8\times10^{-15}$ yr$^{-1}$
(Fischer et al. 2004). At the cosmological time-scale 
$t \sim 10^{10}$ yr\, ($z > 1$), this limit transforms into
$|\Delta \alpha/\alpha| \equiv |(\alpha_z - \alpha)/\alpha| < 3.8\times10^{-5}$,
if $\alpha_z$, the value of $\alpha$ at redshift $z$, is a linear function of $t$. 
The functional dependence of the gauge-coupling constants on $t$
is, however, unknown and theory predicts even oscillations during the course of
the cosmological evolution (e.g., Marciano 1984).
In this regard the astronomical observations are the only way
to test such predictions at different space-time coordinates [see, e.g., 
Mota \& Barrow (2004) where the effects of inhomogeneous space and time evolution
of $\alpha$ are studied].

\begin{table*}[t]
\caption{ESO UVES archive data on the quasar \object{Q 1101--264}}
\label{tbl-1}
\begin{tabular}{rccc r@{$\pm$}l cccccc}
\hline
\noalign{\smallskip}
  & & \multicolumn{6}{c}{QSO} & \multicolumn{4}{c}{ThAr} \\
Exp. & Setting & Date, & Time, & \multicolumn{2}{c}{Seeing,} & T,  & P, &  
Date, & Time, & T,  & P,  \\
No.  &  & y-m-d & h:m:s & \multicolumn{2}{c}{arcsec} & 
$^\circ$C & mb & y-m-d & h:m:s & $^\circ$C & mb \\
(1) & (2) & (3) & (4) & \multicolumn{2}{c}{(5)} &(6) & (7) &(8) & (9) &(10) & (11) \\
\hline
\noalign{\smallskip}
\noalign{\smallskip}
5 & 437 & 2000-02-11 & 04:08:36 & 0.90&0.18& 11.2-11.3 & 992.27-991.87 & 
2000-02-10 & 19:23:55 & 10.9-11.0 & 990.34 \\
6 & 860 & 2000-02-11 & 04:09:47 & \multicolumn{2}{c}{---}& ---  & ---  &
2000-02-10 & 19:25:01 & ---   & ---  \\
\noalign{\smallskip}
11 & 437 & 2000-02-12 & 03:51:57 &0.93&0.11 & 11.4-11.4 & 992.10-991.99 &
2000-02-12 & 13:20:28 & 11.3-11.4 & 993.07 \\
12 & 860 & 2000-02-12 & 03:53:11 &\multicolumn{2}{c}{---} & --- & --- &
2000-02-12 & 13:21:44 & --- & --- \\
\noalign{\smallskip}
19 & 437 & 2000-02-13 & 02:49:57 &0.78&0.10 & 11.4-11.5 & 991.90-991.74 & \\
20 & 860 & 2000-02-13 & 02:51:14 &\multicolumn{2}{c}{---} & --- & --- & \\
\noalign{\smallskip}
21 & 437 & 2000-02-16 & 03:30:15 &0.97&0.32 & 11.2-11.3 & 994.93-994.41 &
2000-02-16 & 13:52:31 & 11.1-11.3 & 994.66 \\
22 & 860 & 2000-02-16 & 03:31:37 &\multicolumn{2}{c}{---} & --- & --- & 
2000-02-16 & 13:53:58 & --- & --- \\
\noalign{\smallskip}
23 & 437 & 2000-02-16 & 04:30:52&0.74&0.01 & 11.2-11.3 & 994.41-994.07 & \\
24 & 860 & 2000-02-16 & 04:32:14&\multicolumn{2}{c}{---} & --- & --- &  \\
\noalign{\smallskip}
\hline
\multicolumn{12}{l}{Cols. 6,7, and 10 list temperature and air pressure inside
the spectrograph at the beginning and the end of the exposure}
\end{tabular}
\end{table*}

The astronomical measurements of the fine-structure splittings
of emission lines in distant galaxies started by Savedoff (1956)
are summarized in a recent comprehensive work by 
Bahcall, Steinhardt \& Schlegel (2004, hereafter BSS): 
$\Delta\alpha/\alpha = (0.7\pm1.4_{\rm stat})\times10^{-4}$ in the
range $0.16 < z < 0.80$\footnote{Statistical errors,
 $\sigma_{\rm stat} \equiv$ dispersion/$\sqrt{n}$,
are used in the text to specify the error of the sample mean, if 
not indicated otherwise.}.
The most stringent bound in the overlapping interval $z \leq 0.45$,
stemming from the radioactive decay rates of certain long-lived nuclei
found in meteoritic data, is set by Olive et al. (2004):
$\Delta\alpha/\alpha = (-8\pm8_{\rm stat})\times10^{-7}$. 
The time interval comparable with the meteoritic analysis is covered by
the Oklo natural reactor ($\Delta t \sim 2\times10^9$ yr). 
Recent reanalysis of the isotopic abundances in the samples taken from Oklo
provides an intriguing result that the value of $\alpha$ was larger in the past:
\daa $\geq 4.5\times10^{-8}$ (Lamoreaux \& Torgerson 2004). 

At higher redshifts the variability of $\alpha$ can be tested by
observations of small shifts between different ionic transitions in the
absorption-line spectra of quasars (Bahcall, Sargent \& Schmidt 1967).
This technique, now known as the alkali doublet (AD) method,
was utilized in numerous studies (for a review, see BSS). 
The best constraint obtained by the AD method is
$\Delta\alpha/\alpha = (-0.5\pm1.5_{\rm stat})\times10^{-5}$ 
(Murphy et al. 2001).

The AD method was generalized by Webb et al. (1999) 
and Dzuba et al. (1999, 2002) 
in the many-multiplet (MM) method which provides an order
of magnitude improvement in the accuracy of the estimations of 
$\Delta\alpha/\alpha$. 
Being applied to 143 metal absorption systems 
(the dominant ions \ion{Mg}{ii}, \ion{Fe}{ii}
at $z < 1.8$, and \ion{Al}{ii}, \ion{Si}{ii} at $z > 1.8$)
identified in the Keck/HIRES
spectra of quasars, the MM method indicates, opposite to the recent Oklo result, 
a decrease of $\alpha$ with cosmic time:
$\Delta\alpha/\alpha = (-5.7\pm1.1_{\rm stat})\times10^{-6}$ in the redshift range
$0.2 < z < 4.2$ (Murphy et al. 2004, hereafter MFWDPW).

A potential concern is that this result includes some systematics.
Indeed, an increasing accuracy of the
$\Delta\alpha/\alpha$ measurements requires a careful consideration of
the intrinsic structure of the atomic transitions which is formed by
the isotope shifts and hyperfine splittings (Levshakov 1994).
For instance, the error of the mean
$\sigma_{\langle\Delta\alpha/\alpha\rangle} = 1.1\times10^{-6}$
corresponds to the error of the line center of $\sim 20$ \ms [see eq.(12)
in Levshakov 2004, hereafter L04] which is about 
40 times smaller than the isotope shift
between $^{26,24}$\ion{Mg}{ii} transitions
$3s \rightarrow 3p_{1/2}, 3p_{3/2}$,
$\Delta v_{24-26} \simeq 850$ \ms
(Drullinger, Wineland \& Bergquist 1980).
Unfortunately, the influence of the isotope shifts cannot be well specified 
since we do not know the isotope abundances at different redshifts.
If the isotope abundance ratio indeed varies with $z$, 
the isotope shifts may imitate the non-zero $\Delta\alpha/\alpha$ value
(Ashenfelter, Mathews, \& Olive 2004; Kozlov et al. 2004, hereafter KKBDF). 
On the other hand,
at metallicities of $Z \sim (0.1-1)\,Z_\odot$, -- typical for the QSO systems
with low ions, -- the isotope abundances
may not differ considerably from terrestrial (Murphy, Webb \& Flambaum 2003; 
Chand et al. 2004, hereafter CSPA).  

The influence of unknown isotopic ratio and of another source of systematics caused by
inhomogeneous ionization structure within the absorber can be considerably
diminished if only one heavy element  like, e.g., \ion{Fe}{ii}, 
is used in the \daa measurement (L04).
In spite of a rather low present accuracy of the theoretical calculations of 
the isotope shift parameters for atoms with more than one valence
electron ($\sim 50$\%, KKBDF),
the isotopic effect for \ion{Fe}{ii} (seven valence electrons in the configuration
$3d^64s$) is less pronounced than that
for \ion{Mg}{ii} (one valence electron in the configuration $3s$) for two reasons:
($i$) iron is heavier and its isotope structure is more compact, 
and ($ii$) the relative abundance of the leading isotope $^{56}$Fe is 
higher\footnote{The terrestrial isotope ratios are
$^{54}$Fe: $^{56}$Fe: $^{57}$Fe: $^{58}$Fe = 5.8 : 91.8 : 2.1 : 0.3, and 
$^{24}$Mg: $^{25}$Mg: $^{26}$Mg = 79 : 10 : 11.}.
Therefore systematic shifts in \daa due to unknown isotopic
compositions should be smaller in
the analysis of the \ion{Fe}{ii} data.

This approach applied to the \ion{Fe}{ii} system identified
at \zabs = 1.15 in the VLT/UVES spectrum of the bright quasar 
\object{HE 0515--4414} ($B = 15.0$) gave
$\Delta\alpha/\alpha = (-0.4\pm1.9_{\rm stat}\pm2.7_{\rm sys})\times10^{-6}$ 
(Quast, Reimers \& Levshakov 2004, hereafter QRL).

The most stringent limit on the variability of $\alpha$,
$\Delta\alpha/\alpha = (-0.6\pm0.6_{\rm stat})\times10^{-6}$, 
standard deviation $\sigma_{\Delta\alpha/\alpha} = 4\times10^{-6}$,
was claimed by CSPA. 
Their result is based on the VLT/UVES observations of 23 absorption systems 
($0.4 \leq z \leq 2.3$) toward 18 QSOs. In this study, the MM method was applied to
a more homogeneous ensemble of \ion{Mg}{ii}, \ion{Si}{ii}, and 
\ion{Fe}{ii} lines which are not strongly saturated and show a less complex structure
as compared with the metal profiles from MFWDPW. 

It should be noted, however, that the measurements with dispersions
$\Delta\alpha/\alpha \sim 2\times10^{-6}$ ($\Delta v \sim 60$ \ms)
are already at the sensitivity limit of the UVES and, hence, they
may be affected by the data reduction procedure and/or by
the changing in time characteristics of the device itself. 
There are three main sources of potential  
systematic errors in the absolute velocity scale:
($i$) temperature and ($ii$) air pressure instability, and ($iii$)
mechanical instabilities of unknown origin. 
For instance, a change of 1 millibar (or a change of
0.3$^\circ$C) induces an error in radial velocities of $\sim 50$ \ms
(Kaufer, D'Odorico \& Kaper 2004). 

In the present paper we investigate step by step the
data reduction procedure usually applied to the VLT/UVES spectra
in order to reveal possible systematics
and to design the optimal measurement technique to avoid them.
The analysis is based on  
another \ion{Fe}{ii} system identified in the absorption-line spectrum of 
quasar \object{Q 1101--264}.

The paper is organized as follows. In Sect.~2 we describe observations and data
reduction performed to construct our \ion{Fe}{ii} sample. Small velocity shifts
(equivalent to 0.1--0.2 pixel size) between the scientific exposures are
analyzed in Sect.~3. Being ignored, such shifts may affect the shapes of the
\ion{Fe}{ii} profiles in the co-added spectra, leading to their inconsistency.
Sect.~4 presents the key procedure used to study the time dependence of $\alpha$,
including error estimation. The results of our analysis are explained in Sect.~5.
Sect.~6 gives the fundamental noise limitation in the \daa measurement due to
photon count and spectral profile. The quality factor similar to that introduced
by Connes (1985) to optimize measurements of the stellar radial velocities is
defined and calculated in this section. 
It can be used to control the sample dispersion in \daa.
The obtained results and future prospects are discussed in Sect.~7. Our conclusions
are given in Sect.~8.

\begin{table}[t]
\centering
\caption{Atomic data of the \ion{Fe}{ii} transitions$^a$, the 
sensitivity coefficients ${\cal Q}^b$, and the isotope mass shift constants
$k_{\rm MS}$$^c$. Estimated errors are given in parentheses}
\label{tbl-2}
\begin{tabular}{c r@{.}l r@{.}l r@{.}l l}
\hline
\noalign{\smallskip}
Mlt. & \multicolumn{2}{c}{$\lambda_{\rm vac}$, } &
\multicolumn{2}{c}{$f$} & \multicolumn{2}{c}{\hspace{-0.5cm}${\cal Q}$} &
$k_{\rm MS}$, \\ 
No.$^d$  &  \multicolumn{2}{c}{\AA} & \multicolumn{2}{c}{ } & 
\multicolumn{2}{c}{ } & \hspace{-0.2cm}cm$^{-1}$\,amu \\
\noalign{\smallskip}
\hline
\noalign{\smallskip}
1u & 2600&1725(1) & 0&23878 & 0&035(4) & --60(20)\\
1u & 2586&6496(1) & 0&06918 & 0&039(4) & --60(20)\\
2u & 2382&7642(1) & 0&320   & 0&035(4) & --63(20)\\
2u & 2374&4603(1) & 0&0313  & 0&038(4) & --63(20)\\
3u & 2344&2130(1) & 0&114   & 0&028(4) & --60(20)\\
7u & 1611&20034(8)& 0&00136 & 0&018(5) & --67(40)\\
8u & 1608&45080(8)& 0&0580  & --0&021(5)& \hspace{0.18cm}67(40)\\
\noalign{\smallskip}
\hline
\multicolumn{8}{l}{$^a$based on the compilation of Murphy et al. (2003);}\\ 
\multicolumn{8}{l}{$^b$defined in Sect.~4; $^c$calculated by Kozlov et al. (2004);}\\
\multicolumn{8}{l}{$^d$multiplet numbers from Morton (2003)}
\end{tabular}
\end{table}

\section{Construction of the \ion{Fe}{ii} sample}

The most accurate measurements of $\Delta\alpha/\alpha$ can be
carried out with unsaturated \ion{Fe}{ii} lines which lie outside
the Ly-$\alpha$ forest and are not corrupted by any other absorption and/or
telluric lines. The components of the \ion{Fe}{ii} lines should be
clearly distinguished.  Besides, a QSO should be a bright object to provide a
high S/N ratio. These requirements are fulfilled for the \zabs = 1.839 system toward
\object{Q 1101--264} with \zem = 2.145 and $V = 16.02$, which was discovered by
Osmer \& Smith (1977). The absorption system was identified by
Carswell et al. (1982) and investigated with high spectral resolution by
Petitjean, Srianand \& Ledoux (2000) and Dessauges-Zavadsky et al. (2003)
who used the same UVES/VLT data as in the present work. 
This system was also studied with lower resolution
in a series of works by Boksenberg \& Snijders (1981),
Young, Sargent \& Boksenberg (1982), Carswell et al. (1984),
and Lanzetta et al. (1987). 

\begin{figure*}[t]
\vspace{0.0cm}
\hspace{-0.3cm}\psfig{figure=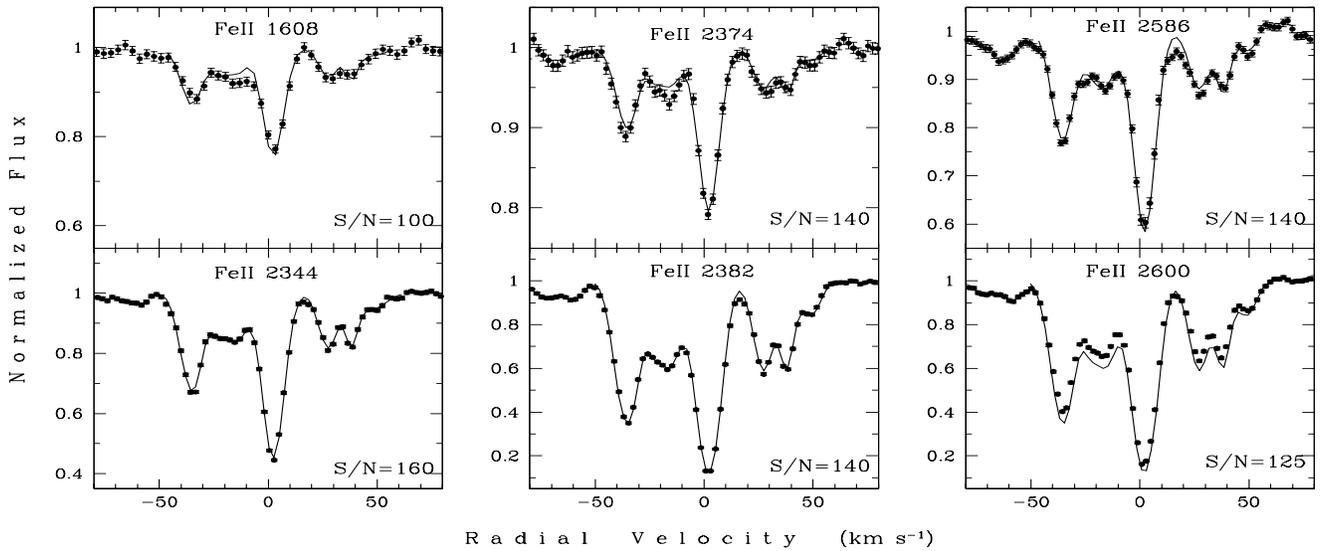,height=15.0cm,width=19.0cm}
\vspace{-7.5cm}
\caption[]{Combined absorption-line spectra of 
\ion{Fe}{ii} associated with the \zabs = 1.839 damped
Ly-$\alpha$ system toward \object{Q 1101--264} (normalized intensities are shown by
dots with 1\,$\sigma$ error bars). The zero radial velocity is fixed at
\zabs = 1.83888. Smooth lines are the synthetic Voigt profiles convolved with the
point-spread spectrograph function. 
The mean signal-to-noise ratio per pixel at the continuum level is indicated in each
panel.  The normalized $\chi^2_{\rm min} =
3.995$ (the number of degrees of freedom $\nu = 272$) shows that the
\ion{Fe}{ii} profiles obtained in the framework of 
the standard data reduction procedure are
not self-consistent
}
\label{fig1}
\end{figure*}

\subsection{Observations and data reduction}

The observations were acquired with the UVES at
the VLT 8.2~m telescope at Paranal, Chile, and the spectral data were
retrieved from the ESO archive. The high resolution spectra
of \object{Q 1101--264} were obtained during the UVES Science Verification programme 
for the study of the Ly-$\alpha$ forest (Kim, Cristiani \& D'Odorico 2002).
The spectra used here were recorded with a dichroic filter which allows
to work with the blue and red UVES arms simultaneously as with two 
independent spectrographs.
The standard settings with central wavelengths at $\lambda$437 nm and $\lambda$860 nm
were used for the blue and red arms, respectively.
From the blue spectra we used only  order 102, and from the red spectra -- orders
90 and 83, where \ion{Fe}{ii} lines suitable for the \daa measurement are observed.
Details of the observations are presented in Table~1. 
Cols.~1 and 2 give the exposure archive number and the setting used.
Cols.~3-7 list the QSO data, and Cols.~8-11 the data of the closest in time 
calibration thorium-argon (ThAr) lamp.

In this work we analyze five exposures (No.~5/6, 11/12, 19/20, 21/22, and 23/24) 
of 3600~s obtained 
over four nights in February 2000 with seeing as indicated in Col.~5 of Table~1.
The slit widths were both set at 0.8 arcsec and the CCDs were read-out in 1$\times$2
binned pixels (spatial$\times$dispersion direction). 
The resulting spectral resolution as measured from the ThAr emission lines is of
FWHM $\simeq 6.0$ \kms\, in the blue ($\lambda \sim 4570$~\AA), 
and of $\simeq 5.4$ \kms\, in the red ($\lambda \sim 7380$~\AA) 

We used the UVES pipeline (the routines implemented in MIDAS-ESO data
reduction package for UVES data) to perform
the bias correction, inter-order background subtraction,
flat-fielding, correction of cosmic rays impacts, 
sky subtraction, extraction of the orders in the pixel space
and wavelength calibration. 
A modified version of the routine `echred' of the context ECHELLE inside
MIDAS was used to calibrate in wavelength the echelle spectra without rebinning.
In this way we have the reduced spectra with their original pixel size in
wavelength. This corresponds to a sampling of
50 m\AA\, (3.3 \kms/pix) for the blue and
55 m\AA\, (2.2 \kms/pix) for the red.
Typical rms of the wavelength calibration is less than one-fiftieth of
a pixel, or $\sigma_{\rm rms} \la 1$~m\AA\, ($\la 60$ \ms).
The observed wavelength scale of each spectrum was transformed into vacuum,
heliocentric wavelength scale (Edl\'en 1966).
At first stage of our analysis 
we followed the standard procedure and added together
the single extracted spectra 
using weights proportional to the square of their S/N.
Before being added, the spectra were re-sampled to an equidistant wavelength grid
(50 m\AA/pix) using linear interpolation. 

\subsection{A test for concordance of the \ion{Fe}{ii} profiles}

The \ion{Fe}{ii} profiles in the \zabs = 1.8389 systems toward \object{Q 1101--264}
consist of at least 6 subcomponents spread over 100 \kms (Dessauges-Zavadsky
et al. 2003). The central component, seen in all but one \ion{Fe}{ii} 
lines from Table~2 (the undetected $\lambda 1611$ \AA\, has the smallest
oscillator strength), is marginally blended in the blue wing with 
weak component at  $v = -17.6$ \kms  (see Fig.~1). 
Henceforth we will use this central component in the
$\Delta\alpha/\alpha$ measurement.

We carefully checked that
the profiles shown in Fig.~1 are free from cosmic rays and telluric absorptions.
The combined spectra reveal high signal-to-noise ratios, $100 \leq$ S/N $\leq 160$.
The continuum levels are very well defined for all iron lines since in all
cases there are large `continuum windows' at both sides of each
\ion{Fe}{ii} absorption. But, nevertheless, 
we failed to find a model which could adequately describe the stacked profiles.
Using the multicomponent Voigt profile fitting procedure 
we obtained the best normalized
$\chi^2_\nu = 3.995$ for the number of degrees of freedom $\nu = 272$. 
The source of such high $\chi^2_\nu$ value is clearly seen in Fig.~1:
some portions of the \ion{Fe}{ii} profiles are not in concordance with each other.

This inconsistency in the combined \ion{Fe}{ii} spectra may reflect some
hidden problems in the standard data reduction procedure.
We investigate this question in the next section.

\section{Velocity shifts between individual exposures}

In order to understand the origin of the discrepancy between \ion{Fe}{ii} profiles,
we firstly investigated distortion and wavelength calibration. 
It turned out that
two \ion{Fe}{ii} lines $\lambda2344$ and $\lambda2374$ lie close to the starting and
end points of the echelle orders where distortions in the spectral sensitivity
are the largest. These lines were excluded from the further analysis.
The remaining \ion{Fe}{ii} lines are observed in the central parts of the echelle
orders. 

Then we selected isolated and unblended ThAr lines from these central spectral
regions and compared their profiles. The time intervals between 
calibration exposures are large and the ThAr spectra were taken under different
conditions. Namely, the differences between the mean temperatures and air pressures 
are 
$\Delta T_{5/6-11/12} = -0.40^\circ$~C, 
$\Delta T_{11/12-21/22} = 0.15^\circ$~C, and
$\Delta P_{5/6-11/12} = -2.73$~mb, 
$\Delta P_{11/12-21/22} = 1.59$~mb, respectively (see Table~1).
Changes in $T$ and $P$ lead to the velocity shifts
between exposures. For instance, emission profiles plotted in
Fig.~2 show that in the blue arm 
$\Delta\lambda_{5-11} \simeq -6$ m\AA\,($\Delta v \simeq -390$ \ms) 
and  $\Delta\lambda_{21-11} \simeq 4$ m\AA\,
($\Delta v \simeq 260$ \ms), 
whereas in the red arm 
$\Delta\lambda_{22-12}$ equals $\Delta\lambda_{21-11}$, as expected, but
$\Delta\lambda_{6-12} \simeq -14$ m\AA\,($\Delta v \simeq -620$ \ms) which
is 1.6 times larger than $\Delta v_{5-11}$ (exposures 11 and 12 are taken
for references). If the velocity drifts were entirely due to 
temperature and pressure changes, then we would observe a monotonous shift 
identical in the blue and red arms, as in the case of the 21st and 22nd exposures. 
The deviation from this behavior revealed for the
5th and 6th exposures indicates that the third factor -- some mechanical 
instabilities -- may also be important. Since it is impossible to control what 
kind of distortions are caused by mechanical instabilities, we remove the 
scientific exposures 5 and 6 from the study.

\begin{figure}[t]
\vspace{0.0cm}
\hspace{0.0cm}\psfig{figure=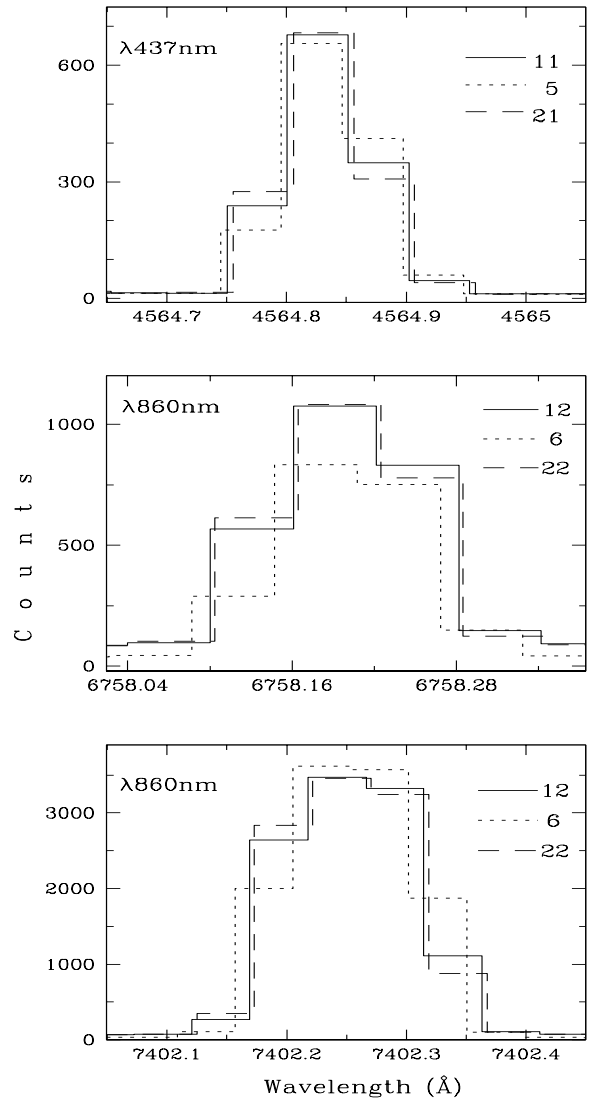,height=15.0cm,width=8.0cm}
\vspace{0.0cm}
\caption[]{Emission line profiles from the thorium-argon calibration
lamp. The lines are placed close to the centers of the UVES echelle orders 
of different settings which are indicated at the top left hand corners 
in the panels. The wavelength step size corresponds to the original pixel width. 
The numbering of the histograms are in accord with the list in Table~1. 
The 11th and 12th exposures are taken as reference. Note the difference
between the relative shifts 5-11 and 6-12, while 21-11 and 22-12 show identical 
velocity offsets
}
\label{fig2}
\end{figure}

This analysis shows that in order to avoid uncertainties in the wavelength calibration
caused by changes in temperature and/or pressure,  
the calibration ThAr lamps must be taken just before and after the scientific
exposure. If for some exposures the relative shifts of the ThAr lines
in the blue and red arms turn out to be different,
these exposures must be excluded from the consideration since they are influenced by
mechanical instabilities which cannot be properly corrected.

Following the standard reduction procedure the remaining scientific exposures
should be co-added to enhance S/N. However, this step can introduce additional
uncertainties because every calibrated exposure has its own velocity offset.
Note that the combination of the exposures
by means of the MIDAS package requires the same starting wavelength,
number of points, and the step size.
Such procedure cancels out the original non-zero
offsets between the individual exposures and smears out small shifts.
This can perturb the line centroids, especially those of narrow absorption
lines with sharp intensity gradients.
Since in the \daa measurement the accuracy of the line position is the key factor,
it is more appropriate to work with individual exposures which have lower S/N rather
than with a high S/N combined spectrum.

Accounting for these arguments we selected the scientific exposures with
\ion{Fe}{ii} lines shown in Fig.~3. All spectra are vacuum and heliocentric
calibrated with only one ThAr spectrum (exposures 11 and 12 for the blue and 
red frames, respectively), they are not combined and not resampled.
The unknown velocity offsets arising from such calibration 
are canceled out in the differential measurements of \daa\, as
described in the next section.

The key line in our approach is \ion{Fe}{ii} $\lambda1608$ since its sensitivity
coefficient is negative, while the other iron transitions 
have positive ${\cal Q}$-values (Table~2). A relatively low
strength of the $\lambda1608$ line requires S/N $\ga 30$ in individual exposures
in order to measure its center accurately. This requirement is fulfilled for 
\object{Q 1101--264} (see Fig.~4, where the S/N values are indicated). 
It is also important that for this system the accuracy of the
normalization of the \ion{Fe}{ii} profiles leaves
no room for doubts since the local 
continua (the horizontal lines in Fig.~3) are calculated with high precision.
In each panel in Fig.~3, dots with error bars indicate the mean
intensities and their 1$\sigma$ uncertainties utilized in the
linear regression analysis applied to determine
the local continuum level. The uncertainty of the calculated continuum 
is found to be less than 1\%\, for all selected \ion{Fe}{ii} spectra.

\begin{figure*}[t]
\vspace{0.0cm}
\hspace{0.0cm}\psfig{figure=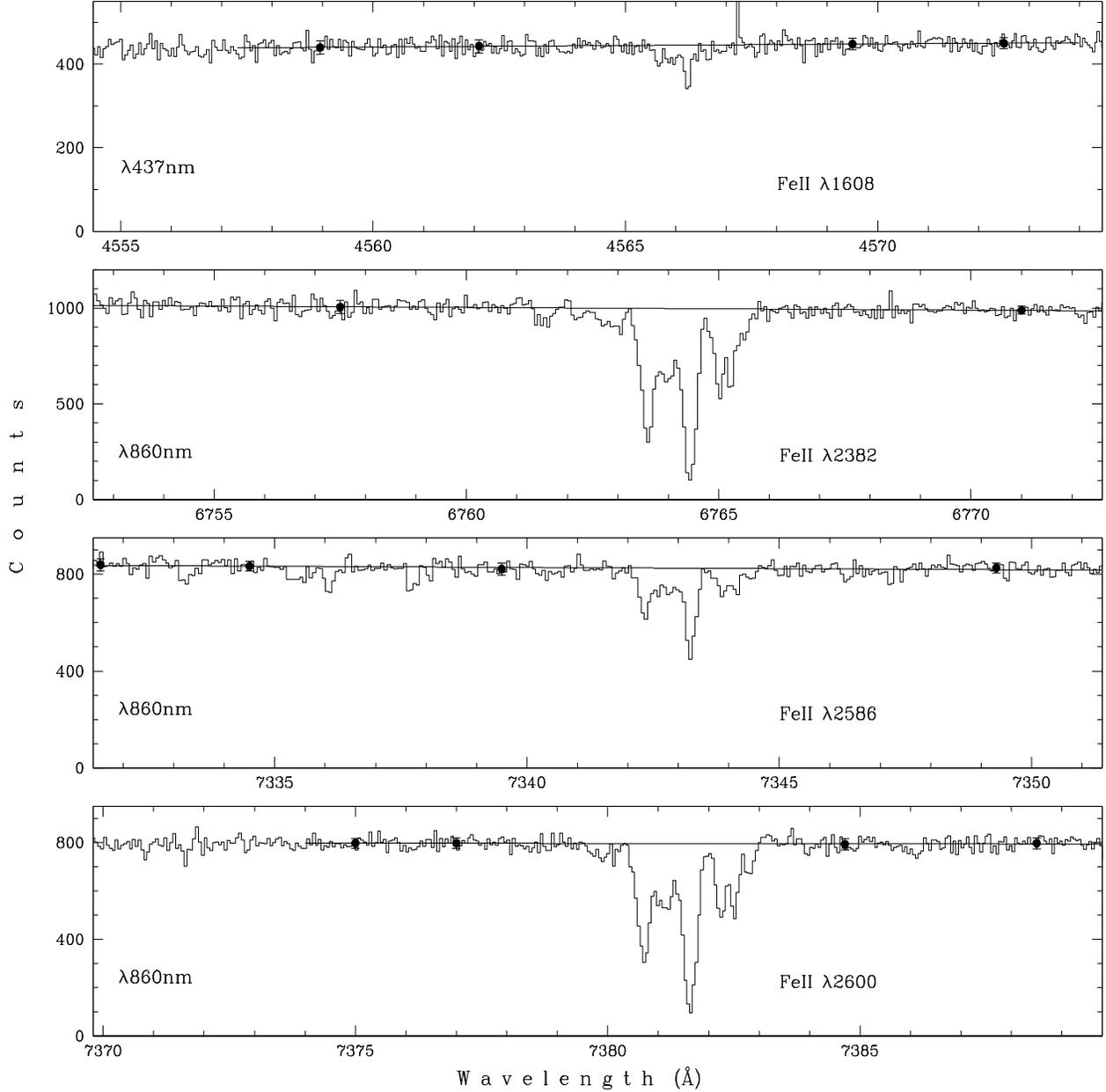,height=19.0cm,width=14.0cm}
\vspace{-2.0cm}
\caption[]{Unnormalized portions of the \object{Q 1101--264} 
spectra with the \ion{Fe}{ii}
lines (\zabs = 1.839) selected for the $\Delta\alpha/\alpha$ measurement.
All lines are located near the centers of the UVES echelle orders. 
The corresponding settings are indicated
at the bottom left hand corners in each panel. Dots with error bars are the mean
intensities and their 1$\sigma$ uncertainties -- the values
used to estimate the local continuum level by means of the linear regression analysis.
For the selected \ion{Fe}{ii} lines, the continuum level is known with an 
accuracy better than 1\%
}
\label{fig3}
\end{figure*}

\section{The single ion differential $\alpha$ measurement}

The standard many-multilet (MM) technique (Webb et al. 1999,
Dzuba et al. 1999, 2002) calculates \daa\, using a quite complex procedure. 
This leads to potential systematic uncertainties
which may affect both the line profile and the origin of the wavelength scale.
The systematic effects are thoroughly discussed in 
Murphy et al. (2003), BSS, and L04. In this section we describe how to 
use a modified MM procedure to calculate \daa\, {\it directly} from the 
differences between the wavelengths of a pair of \ion{Fe}{ii} transitions
observed in the individual exposures. Thereafter this approach is called
a `single ion differential $\alpha$ measurement' (SIDAM).
An important point is that being applied to the lines from the {\it same}
exposure, this method does not depend
on the {\it unknown offsets of the wavelength scale}.

The MM method utilizes the fact that the energy of each line
transitions depends individually on a change in $\alpha$. Namely,
the frequency of each transition has its own relativistic correction
to the changes in $\alpha$ which is
expressed by the coefficient $q$ (Dzuba et al. 1999, 2002). Then,
the systemic rest wavenumber $\omega_z = 1/\lambda_z$ is given by
\begin{equation}
\omega_z = \omega_0 + q\,(\Delta\alpha/\alpha)\,
(2 + \Delta\alpha/\alpha)\; ,
\label{eq:A1}
\end{equation}
where $\omega_0 = 1/\lambda_0$ is the laboratory wavenumber.

In fact, this approach is similar to the method developed by
Varshalovich \& Levshakov (1993) in order to infer the cosmological
variability of the proton-electron mass ratio, $\mu = m_{\rm p}/m_{\rm e}$, 
from the analysis of molecular hydrogen H$_2$ absorption lines.
The dependence of the frequencies of electron-vibro-rotational
transitions on a change in $\mu$ differs for individual transitions and
can be characterized by the dimensionless sensitivity coefficient ${\cal K}$.
The estimation of $\Delta\mu/\mu$ can be obtained from linear
regression analysis of the $\{z,{\cal K}\}$ pairs from a sample of the
H$_2$ lines (Potekhin et al. 1998; Levshakov et al. 2002b).
Below this technique is applied to the \daa measurement.

We can re-write (\ref{eq:A1})
in linear approximation ($|\Delta\alpha/\alpha|\ll1$) in the form (L04): 
\begin{equation}
z_i = z_0 + \kappa_\alpha\,{\cal Q}_i\; ,
\label{eq:A2}
\end{equation}
where $z_i = \lambda_{{\rm obs},i}/\lambda_{0,i} - 1$, 
${\cal Q}_i = q_i/\omega_{0,i}$ the
dimensionless sensitivity coefficient,
and the slope parameter $\kappa_\alpha$ is given by
\begin{equation}
\kappa_\alpha = - 2(1 + z_0)(\Delta \alpha/\alpha)\; .
\label{eq:A3}
\end{equation}
If $\Delta \alpha/\alpha$ is non-zero, $z_i$ and ${\cal Q}_i$ will be correlated.
The slope $\kappa_\alpha$ and the intercept $z_0$ can be
found from the linear regression analysis of the observed redshifts of
the line centroids $z_i$ vs. ${\cal Q}_i$ (for more details, see
L04 and  QRL, where this method is called a `regression' MM analysis, RMM).

In this approach, the value of $\Delta\alpha/\alpha$
can be directly estimated from a pair of lines with different
sensitivity coefficients. It is easy to show that
\begin{equation}
\frac{\Delta\alpha}{\alpha} = \frac{(z_2 - z_1)}
{(2 + z_1 + z_2)({\cal Q}_1 - {\cal Q}_2)}\; ,
\label{eq:A4}
\end{equation}
or
\begin{equation}
\frac{\Delta\alpha}{\alpha} = \frac{(z_2 - z_1)}
{2(1 + \bar{z})({\cal Q}_1 - {\cal Q}_2)}\; .
\label{eq:A5}
\end{equation}
The ratio $(z_2 - z_1)/(1 + \bar{z})$ is invariant and does not depend on
systematic velocity shifts between different scientific exposures.

The computational procedure can be 
slightly modified if we use another variables similar to those
introduced in the AD analysis by BSS.

Let us consider an absorption line system at redshift $z$ where a set of
\ion{Fe}{ii} transitions is formed at the corresponding cosmic time $t$.
Then, the ratio of the observed wavelengths $\lambda_i(t)$ to $(1+z)$ defines
the rest wavelengths $\lambda'_i(t)$ which may differ from their present-day
values $\lambda_i(0)$, if $\Delta\alpha/\alpha \neq 0$.
From (\ref{eq:A1}) one obtains in linear approximation
\begin{equation}
\lambda'_i(t) = \lambda_i(0)\,(1 - 2{\cal Q}_i\,\Delta\alpha/\alpha)\; .
\label{eq:A6}
\end{equation}

Following BSS, we define $R(t)$ and $\eta$ by the relations 
\begin{equation}
R(t) = \frac{\lambda_2(t) - \lambda_1(t)}{\lambda_1(t) + \lambda_2(t)}\; ,
\label{eq:A7}
\end{equation}
and
\begin{equation}
\eta = \frac{\lambda_2}{\lambda_1} - 1\; .
\label{eq:A8}
\end{equation}

The cosmological redshift of the absorber 
in the expression for $R$ cancels out and thus 
\begin{equation}
\frac{\lambda_2(t) - \lambda_1(t)}{\lambda_1(t) + \lambda_2(t)} =
\frac{\lambda'_2(t) - \lambda'_1(t)}{\lambda'_1(t) + \lambda'_2(t)}\; . 
\label{eq:A9}
\end{equation}
From (\ref{eq:A6})-(\ref{eq:A9}), one obtains after some algebra
[taking into account $\eta(t) \approx \eta(0)$]
\begin{equation}
\frac{\Delta\alpha}{\alpha} = \frac{1}{4({\cal Q}_2 - {\cal Q}_1)}\,
\frac{\eta (\eta + 2)}{\eta + 1}\,
\left(1 - \frac{R(t)}{R(0)}\right)\; .
\label{eq:A10}
\end{equation}
In some cases (see Sect.~7.2) 
this formulation can be more convenient for calculations.

\begin{figure*}[t]
\vspace{0.0cm}
\hspace{0.0cm}\psfig{figure=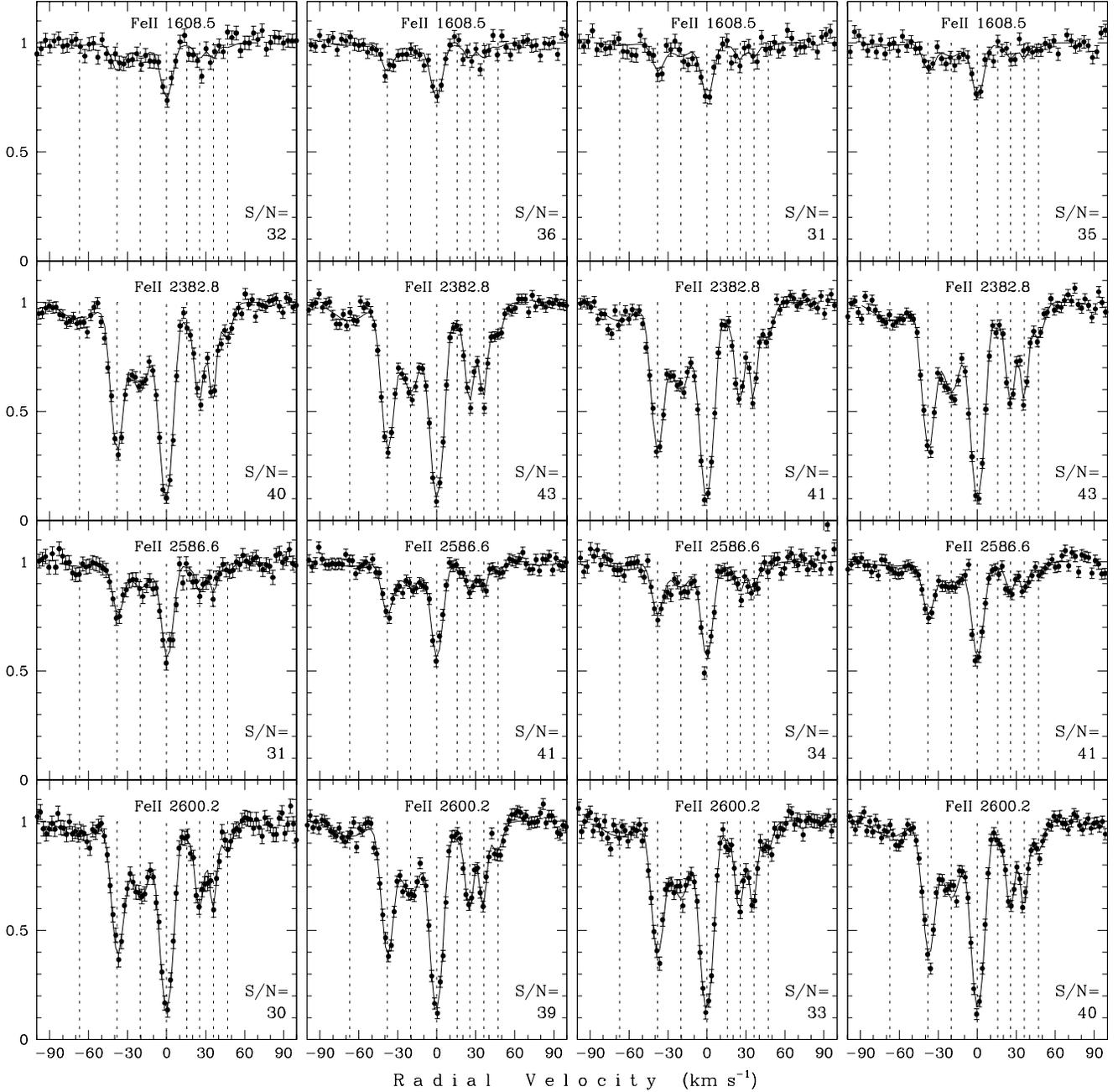,height=18.0cm,width=16.0cm}
\vspace{-0.5cm}
\caption[]{Individual exposures of the \ion{Fe}{ii} lines in the spectrum 
of \object{Q 1101--264} (normalized intensities are shown by
dots with 1$\sigma$ error bars) and over-potted
synthetic profiles (smooth curves) calculated from the joint analysis. 
The minimization procedure gives $\chi^2_{\rm min} = 1.097$ per degree of
freedom (the number of data points $m = 973$, the number of fitting 
parameters $p = 39$).
Dashed vertical lines mark positions of the individual components.
The mean signal-to-noise ratio per pixel at the continuum level is indicated in each
panel.  
Columns from the left to the right correspond to the exposures No. 11/12, 19/20,
21/22, and 23/24, in accord with Table~1. Spectral data shown in these columns
are centered relative to the following fixed redshifts (from the left to the right):
1.8389040, 1.8389041, 1.8389065, and 1.8389009
}
\label{fig4}
\end{figure*}

\subsection{Error estimations in the SIDAM technique} 

In this section we calculate the uncertainty on individual 
$\Delta\alpha/\alpha$ measurement caused by the errors in the wavelengths.
The standard method of error propagation is used. 

From eq.(\ref{eq:A4})
we deduce (taking into account $z_2 \approx z_1$)
\begin{equation}
\sigma^2_{\Delta\alpha/\alpha} =
\frac{1}{4({\cal Q}_1 - {\cal Q}_2)^2}\,\left(\delta^2_{z_1} +
\delta^2_{z_2}\right)\; ,
\label{eq:A11}
\end{equation}
where the relative error $\delta_z = \sigma_z/(1+z)$ is equal to
\begin{equation}
\delta_{z} = (\delta^2_{\lambda} + \delta^2_{\lambda_{0}})^{1/2}\; .
\label{eq:A12}
\end{equation}

If $\delta_{\lambda_0} \ll \delta_\lambda$ and
$\delta_{\lambda_1} \approx \delta_{\lambda_2}$, then
\begin{equation}
\sigma_{\Delta\alpha/\alpha} = \frac{1}{\sqrt{2}\,|{\cal Q}_1 - {\cal Q}_2|}\,
\delta_\lambda\; .
\label{eq:A13}
\end{equation}
If the relative errors for both observational and laboratory lines are
comparable, then the best estimation of $\Delta\alpha/\alpha$ is known with
the error
\begin{equation}
\sigma_{\Delta\alpha/\alpha} = \frac{1}{|{\cal Q}_1 - {\cal Q}_2|}\,
\delta_\lambda\; .
\label{eq:A14}
\end{equation}

Using relation (\ref{eq:A14}), it is easy to estimate the highest 
possible precision in the $\Delta\alpha/\alpha$ measurement with the laboratory
\ion{Fe}{ii} data at hand. Assuming that the observed line centers are known
with the accuracy
$\delta_\lambda \simeq 5\times10^{-8}$ (see Table~2), one obtains
$\tilde{\sigma}_{\Delta\alpha/\alpha} \simeq 8\times10^{-7}$. 
This corresponds to the uncertainty $\sigma_\lambda = 0.25$ m\AA\,
($\Delta v = 15$ \ms) at 5000 \AA.
The value $\Delta v \sim 15$ \ms is 4 times lower than the limit sensitivity
of the UVES, and, hence, with good statistics 
(the sample size $\ga 16$) \daa can, in principle, be probed on the level of 
$\la 10^{-6}$. 

We note that the best precision reached today in the measurements of
stellar radial velocities of relatively bright stars is a few \ms.
Observations with HARPS provide, for example, the rms uncertainty of
2 \ms (Pepe et al. 2004) which is close to the HARPS fundamental noise limitation.
For the $\Delta\alpha/\alpha$ measurement with UVES, the fundamental noise 
limitation due to spectral profile and photon count is considered in Sect.~6.

\begin{table*}[t]
\centering
\caption{SIDAM analysis: optimized centroid positions of the
\ion{Fe}{ii} lines in the main subcomponent and $\Delta\alpha/\alpha$
calculated with (\ref{eq:A4}) or (\ref{eq:A10})
}
\label{tbl-3}
\begin{tabular}{ccccccc}
\hline
\noalign{\smallskip}
Exp.  & $\lambda_{\rm obs}$, \AA & $\lambda_{\rm obs}$, \AA &
$\lambda_{\rm obs}$, \AA & $\lambda_{\rm obs}$, \AA & 
\multicolumn{1}{c}{$(\Delta\alpha/\alpha)$}& pair, \\
No. & $\lambda1608$ & $\lambda2382$ & $\lambda2586$ & $\lambda2600$ & 
\multicolumn{1}{c}{(in units of $10^{-5}$)} & $\lambda_1/\lambda_2$ \\
(1) & (2) & (3) & (4) & (5) & \multicolumn{1}{c}{(6)} & (7) \\
\noalign{\smallskip}
\hline
\noalign{\smallskip}
11/12 & 4566.24408 & 6764.43913 & 7343.27915 & 7381.65008 & 1.199 & 1608/2382\\
      &  &  &  &  &  \hspace{-0.15cm}--2.103 & 1608/2586\\
      &  &  &  &  & 0.098 & 1608/2600 \\
\noalign{\smallskip}
19/20 & 4566.23755 & 6764.44738 & 7343.26356 & 7381.64359& 
\hspace{-0.15cm}--1.047 &1608/2382 \\
      & & & & & \hspace{-0.15cm}--1.525& 1608/2586 \\
      & & & & &  \hspace{-0.15cm}--0.393& 1608/2600 \\
\noalign{\smallskip}
21/22 & 4566.24661 & 6764.43913 & 7343.25994 & 7381.63961&1.668& 1608/2382\\
      & & & & & 0.539& 1608/2586\\
      & & & & & 1.860& 1608/2600  \\
\noalign{\smallskip}
23/24 & 4566.25102 & 6764.44933 & 7343.27203 & 7381.65019&1.209& 1608/2382 \\
      & & & & &  \hspace{-0.15cm}--0.028 &  1608/2586 \\
      & & & & & 1.442 &  1608/2600 \\
\noalign{\medskip}
\multicolumn{5}{r}{sample mean $\langle \Delta\alpha/\alpha\rangle$:} & 
\multicolumn{1}{c}{0.24} &  \\
\multicolumn{5}{r}{error of the mean $\sigma/\sqrt{n}$:} & 
\multicolumn{1}{c}{0.38} & \\
\noalign{\smallskip}
\noalign{\smallskip}
\hline
\end{tabular}
\end{table*}

\subsection{Errors caused by the isotopic shifts}

The \daa measurement on the scale of $\sim 8\times10^{-7}$
can be affected by the isotopic shifts if the ratio 
${\cal I} =\, ^{54}$Fe/$^{56}$Fe
varies between the absorbers (the influence of the isotope $^{57}$Fe is negligible
because of its relatively low abundance).
Following KKBDF and using their 
mass shift constants $k_{\rm MS}$ for \ion{Fe}{ii} transitions (Table~2),
we can estimate the shift of the line center of gravity  
mimicking a non-zero \daa:
\begin{equation}
\delta\omega_c = \frac{x}{100}\,\Delta\omega^{A',A} \approx  \frac{x}{100}\,
k_{\rm MS}\left(\frac{1}{A'} - \frac{1}{A}\right)
\; ,
\label{eq:A16a}
\end{equation}
where $A'$ and $A$ are the isotope mass numbers, and $x$ (in per cents) 
shows the reduced fraction of the leading isotope $A$. 
If, for example, $x = 10$\% (i.e. ${\cal I}_z/{\cal I}_\odot \simeq 3$), 
then $ \delta\omega_c \simeq -0.004$ cm$^{-1}$ for
\ion{Fe}{ii} transitions (see Table~1 in KKBDF).

Numerical simulations of the explosive yields show, however, that
at low metallicities $Z < Z_\odot$ the ratio ${\cal I}_z/{\cal I}_\odot \la 1$ 
(e.g., Chieffi \& Limongi 2004). Thus, in the high-$z$ absorbers 
one may expect an approximately constant
isotopic shift $\delta\omega_c = -0.12\,k_{\rm MS}/A^2 \simeq 0.002$ cm$^{-1}$,
which is equivalent to the positive shift \daa $\simeq 7.7\times10^{-7}$ 
(note that an offset of the velocity scale of 10-15 \ms can produce the same effect).
However, this shift is canceled out in
the differential \daa measurements of a few high-$z$ absorbers having
the same metallicities.  This can essentially  
improve the limiting accuracy of 
$\sigma_{\Delta\alpha/\alpha} \sim 10^{-6}$ set by KKBDF for 
\ion{Fe}{ii} samples.

\section{Analysis and the \daa\, results}

In this section we study the \ion{Fe}{ii} profiles selected from
individual exposures and derive the position of the
line centroid of the main absorption component seen at zero radial velocity
in Fig.~4. The total number of the analyzed profiles is $L = 16$. 
Since the line profiles are very complex, the  
distribution of the velocity components along the line of sight is crucial
for the following $\Delta\alpha/\alpha$ measurement. To construct the model
for the radial velocity distribution, we start with the analysis of the profiles from
the individual scientific exposures to fix their reference frames -- the mean
\zabs values. The following redshifts were determined:
$z_{11/12} = 1.8389040$, $z_{19/20} = 1.8389041$,
$z_{21/22} = 1.8389065$, and  $z_{23/24} = 1.8389009$.

Our model is based on the natural assumption that \ion{Fe}{ii} lines have similar
profiles, i.e., 
(1) the number of subcomponents $n_{\rm s}$ is identical for all \ion{Fe}{ii} lines,
(2) the Doppler $b_i$ parameters are identical for the same
$i$th subcomponents, (3) the relative intensities of the subcomponents
$r_{i,j}$ and (4) the relative radial velocity differences
$\Delta v_{i,j}$ between the subcomponents are fixed for the absorber. 
We also assume that the main broadening is caused by bulk motion.

Then, the \ion{Fe}{ii} profile is described by the sum of $n_{\rm s}$
Voigt functions:
\begin{equation}
\tau^{(\ell)}_v = N_1\,\sum^{n_{\rm s}}_{i=1}\,r_{i,1}\,{\cal V}\left[
(v - v_\ell - \Delta v_{i,1})/{b_i} \right]\; ,
\label{eq:E6}
\end{equation}
where $\tau^{(\ell)}_v$ is the optical depth at radial velocity $v$ within the line
$\ell$, $N_1$ is the column density of the main component, 
$r_{i,1} = N_i/N_1$, 
$v_\ell$ is the center of the main component in the line $\ell$,
and $\Delta v_{1,1} = 0$. 

Our model is fully defined by specifying $N_1$, 
$\{b_i\}^{n_{\rm s}}_{i=1}$,
$\{\Delta v_{i,1}\}^{n_{\rm s}-1}_{i=1}$,
$\{r_{i,1}\}^{n_{\rm s}-1}_{i=1}$, and
$\{v_\ell\}^{L}_{\ell=1}$.
All these parameters are components of the parameter vector
$\theta = \{\theta_1, \theta_2, \ldots, \theta_p\}$.
To estimate $\theta$ from the \ion{Fe}{ii} profiles, we minimize the objective
function
\begin{equation}
\chi^2(\theta) = \frac{1}{\nu}\,\sum^{L}_{\ell=1}\,\sum^{m_\ell}_{j=1}\,
\left[ {\cal F}^{\rm cal}_{\ell,j}(\theta) - {\cal F}^{\rm obs}_{\ell,j}
\right]^2/\sigma^2_{\ell,j}\; ,
\label{eq:E7}
\end{equation}
where ${\cal F}^{\rm obs}_{\ell,j}$ is the observed normalized intensity of the
spectral line $\ell$, $\sigma_{\ell, j}$ is the experimental error within the
$j$th pixel of the line profile, 
and $\nu = \left( \sum^L_\ell\,m_\ell\, -\,p \right)$ 
is the number of degrees of freedom.
${\cal F}^{\rm cal}_{\ell,j}(\theta)$ is the
calculated intensity convolved with the spectrograph point-spread function.

\begin{figure}[t]
\vspace{0.0cm}
\hspace{0.0cm}\psfig{figure=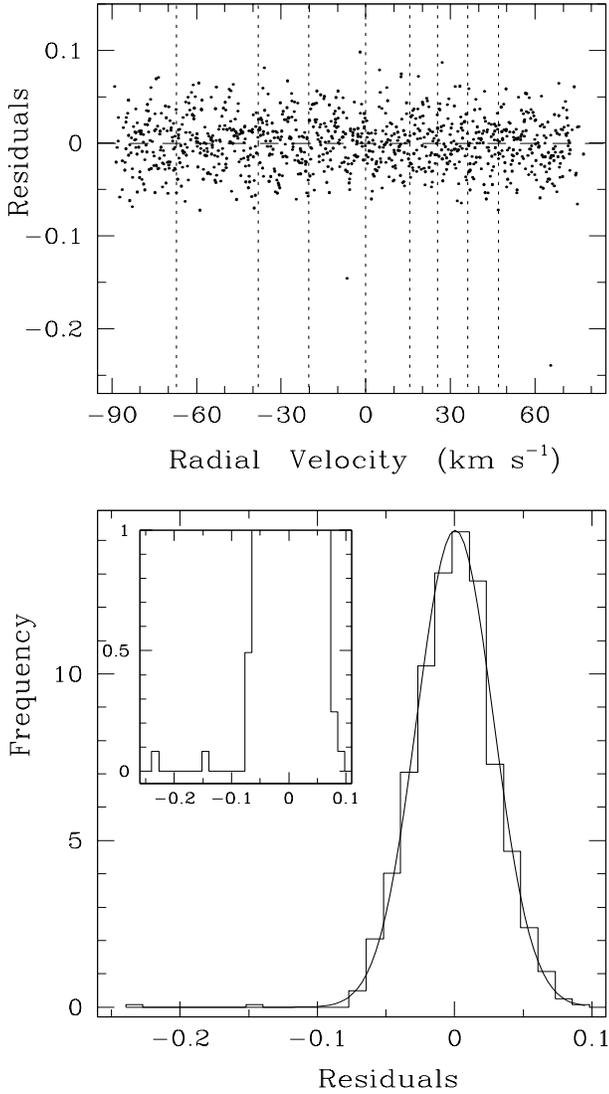,height=17.0cm,width=9.0cm}
\vspace{-2.5cm}
\caption[]{{\it Upper panel}: The residuals $\epsilon =
({\cal F}^{\rm cal} - {\cal F}^{\rm obs})$ for all \ion{Fe}{ii} profiles
from Fig.~4. Dashed vertical lines indicate the positions of the subcomponents.
{\it Lower panel}: Shown are the distribution of $\epsilon$ (histogram) and the
overplotted normal distribution (smooth curve) with the sample mean
$\langle \epsilon \rangle = 3.5\times10^{-4}$ and the dispersion
$\sigma_\epsilon = 2.9\times10^{-2}$ estimated from the total data set
except for two `hot pixels' with $\epsilon < -0.1$ (seen in the inset)
}
\label{fig5}
\end{figure}

The synthetic profiles for the best $\chi^2$
are shown by the smooth lines in Fig.~4.
The optimal number of subcomponents is $n_{\rm s} = 8$. Their positions
are marked by vertical dotted lines in the panels in Fig.~4.
The corresponding $\chi^2_{\rm min} = 1.21$ (for $m = 975$ and $p = 39$)
is, however, too high: 
with $\nu = 936$, the expected mean is $\chi^2_\nu = 1\pm0.05$ ($1\sigma$ c.l.). 
However, the analysis of the residuals $\epsilon =
({\cal F}^{\rm cal} - {\cal F}^{\rm obs})$ shown in Fig.~5 reveals two `hot pixels'
with $\epsilon < -0.1$ which deteriorate this $\chi^2_{\rm min}$ 
value\footnote{The first `hot pixel' with $\epsilon = -0.15$
is seen in Fig.~4, exposures 23/24, panel \ion{Fe}{ii} $\lambda2586.6$, at
$v = -66.6$ \kms, whereas the second one with $\epsilon = -0.24$, the same
exposures, panel \ion{Fe}{ii} $\lambda1608.5$, 
$v = 65.6$ \kms is beyond the frame.}.
After removing these points, we find $\chi^2_{\rm min} = 1.097$ (for $m = 973$ and
$p = 39$).

The estimated wavelengths of the main \ion{Fe}{ii} components 
are listed in Table~3. These are the best fitting quantities. We do not
calculate their errors since the spectral data from different exposures show
very similar S/N ratios which allow us to calculate the sample mean
$\langle \Delta\alpha/\alpha \rangle$ 
without weights. The results are given in Col.~6 of Table~3. 
Col.~7 lists the \ion{Fe}{ii} pairs used in a particular 
$\Delta\alpha/\alpha$ measurement. We find
$\langle \Delta\alpha/\alpha \rangle = (2.4\pm3.8)\times10^{-6}$,
and the corresponding root mean square 
$\sigma_{\rm rms} = 1.3\times10^{-5}$.

\section{The fundamental noise limitation in the $\Delta\alpha/\alpha$ measurement}

In Sect.~4, we described the computational procedure used in this study --
the single ion differential $\alpha$ measurement, SIDAM.
A simple form of the deduced eqs.~(\ref{eq:A4}) and (\ref{eq:A10})
allows us to compute directly the fundamental
uncertainty in the $\Delta\alpha/\alpha$ measurement due to photon noise.
This analysis is based on the results obtained by Connes (1985)
and by Bouchy, Pepe \& Queloz (2001) who calculated the fundamental noise
limitation in the Doppler shift measurements.

Let us consider a digitalized and calibrated spectra of a pair of \ion{Fe}{ii}
lines which are obtained with a high stability spectrograph. 
Let $\Delta\lambda_{\rm pix}$ be the pixel size (the wavelength interval between 
pixels). Assume further 
that the spectrograph point-spread function can be described
by a Gaussian with FWHM$_{\rm sp} = 2\Delta\lambda_{\rm pix}$ (the Nyquist
limit). The observed \ion{Fe}{ii} lines are supposed to be isolated, 
single component, and unsaturated.
Their apparent width, FWHM$_{\rm line}$, is caused by the convolution of the 
spectrograph point-spread function with the `true' profile. The width of the 
true profile is defined by the quadratic sum of the thermal and turbulent components:
FWHM$^2_{\rm true}$ = FWHM$^2_{\rm therm}$ + FWHM$^2_{\rm turb}$.  
Since \ion{Fe}{ii} lines are usually observed in the damped Ly-$\alpha$ systems
where kinetic temperature is low ($\sim 100$~K, FWHM$_{\rm therm} \sim 0.3$ \kms)
and the turbulent broadening is a few \kms, 
FWHM$_{\rm true}$ might be less or about FWHM$_{\rm sp}$. 

The error in the line center caused by counting statistics is given by 
Bohlin et al. [1983, eq.(A15)]:
\begin{equation}
\sigma_\lambda = \frac{\Delta\lambda_{\rm pix}}{W_{\rm obs}}\,
\frac{1}{\sqrt{{\cal N}_{\rm e}}}\,\frac{M\sqrt{M}}{\sqrt{12}}\,
\Delta\lambda_{\rm pix}\; ,
\label{eq:B1}
\end{equation}
where $W_{\rm obs}$ is the observed equivalent width, 
${\cal N}_{\rm e}$ is the mean number of photoelectrons per pixel at the
continuum level, and $M$ is the number of pixels covering the line
profile\footnote{The second term in Bohlin's et al. eq.(A15) is neglected 
since we assume the lines are rather strong (the apparent central
optical depth $\tau_0 \sim 1$), and isolated, i.e. the local continuum level  
is known with a sufficiently high precision.}.

For Gaussian profiles, $M$ can be equal to 
$2.5$\,FWHM$_{\rm line}/\Delta\lambda_{\rm pix}$, producing
\begin{equation}
M = 5\xi\; ,
\label{eq:B2}
\end{equation}
where 
$\xi = [1 + ({\rm FWHM}_{\rm true}/{\rm FWHM}_{\rm sp})^2\,]^{1/2}$.

By analogy with the Connes procedure, we can characterize a line profile
by a dimensionless quality factor $Q$ which is independent on the flux:
\begin{equation}
Q = \frac{W_{\rm obs}}{\Delta\lambda_{\rm pix}}\,
\frac{\sqrt{12}}{5\sqrt{5}\,\xi\sqrt{\xi}}\; ,
\label{eq:B3}
\end{equation}
and re-write (\ref{eq:B1}) in the form
\begin{equation}
\sigma_\lambda = \frac{\Delta\lambda_{\rm pix}}{Q\,\sqrt{{\cal N}_{\rm e}}}\; .
\label{eq:B4}
\end{equation}
This equation shows that the error $\sigma_\lambda$ decreases almost quadratically
with decreasing wavelength bin per pixel
($W_{\rm obs}$ does not depend on the spectral
resolution, and $\xi \sim \sqrt{2}$, if 
FWHM$_{\rm sp} \sim$ FWHM$_{\rm true}$) under the assumption that $N_{\rm e}$ is
keeping constant.
The quality factor for \ion{Fe}{ii} lines from the \zabs = 1.839
system ranges between 0.1 and 0.7 (Table~4).   

The total number of photoelectrons can be estimated from the specific
flux $J^V_\nu = 3.81\times10^{-20}$ ergs s$^{-1}$ cm$^{-2}$ Hz$^{-1}$  
of a $mv = 0$ star outside the Earth's atmosphere\footnote{For $B$ and $R$ filters,
the corresponding intensities are $F^B_\ast = 1.52\times10^3$ and
$F^R_\ast = 0.65\times10^3$ photon cm$^{-2}$ s$^{-1}$ \AA$^{-1}$.}:
$F^V_\ast = 1.05\times10^3$ photon cm$^{-2}$ s$^{-1}$ \AA$^{-1}$. 
Then ${\cal N}_{\rm e}$ is given by:
\begin{equation}
{\cal N}_{\rm e} = \frac{F^V_\ast\,s_{\rm tel}\,\varepsilon_{\rm tot}\,
t_{\rm exp}\,\Delta\lambda_{\rm pix}}{10^{0.4\,mv}}\; .
\label{eq:B5}
\end{equation}
Here $s_{\rm tel}$ is the telescope area in cm$^2$; $\varepsilon_{\rm tot}$
the overall detection efficiency of the telescope, spectrograph and
detector, corrected for the contribution of the
atmosphere; $t_{\rm exp}$ the exposure time in s; $mv$ the visual magnitude
of the quasar.

\begin{table}[t]
\centering
\caption{Fundamental photon noise limits 
$\tilde{\sigma}^{\rm lim}_{\Delta\alpha/\alpha}$ 
(in units of $10^{-5}$)  
for 
\ion{Fe}{ii} pairs of lines ($\lambda1608/X$) from the \zabs = 1.839 system toward
\object{Q 1101--264}. The exposure time $t_{\rm exp}$ and the total efficiency
$\varepsilon_{\rm tot}$ are set to 3600~s and 0.15, respectively}
\label{tbl-4}
\begin{tabular}{lr@{.}l r@{.}l r@{.}l r@{.}l}
\hline
\noalign{\smallskip}
\multicolumn{1}{r}{Line:}  & \multicolumn{2}{c}{$\lambda1608$} & 
\multicolumn{2}{c}{$\lambda2382$} & 
\multicolumn{2}{c}{$\lambda2586$} & 
\multicolumn{2}{c}{$\lambda2600$} \\
\noalign{\smallskip}
\hline
\noalign{\smallskip}
$W_{\rm obs}$, \AA & 0&0446 & 0&2779 & 0&1239 & 0&2770 \\
$\Delta\lambda_{\rm pix}$, \AA & 0&05 & 0&06 & 0&05 & 0&05 \\
$\xi$ & 1&83 & 1&83 & 1&93 & 1&93 \\
$Q$-factor & 0&11 & 0&58 & 0&29 & 0&64 \\
\noalign{\smallskip}
$\sigma^{\rm lim}_\lambda$, m\AA & 5&7 & 1&3 & 2&3 & 1&0 \\
\noalign{\smallskip}
$\tilde{\sigma}^{\rm lim}_{\Delta\alpha/\alpha}$ & \multicolumn{2}{l}{---} &
1&1 & 1&0 & 1&1 \\
\noalign{\smallskip}
\hline
\end{tabular}
\end{table}

At 5500~\AA, the UVES efficiency $\varepsilon_{\rm tot} \simeq 0.15$
(Kaufer et al. 2004), and with a  3600~s exposure (${\cal N}_{\rm e} \simeq 5670$),
one expects a fundamental uncertainty of about 1-5 m\AA\, in $\sigma_\lambda$
for a $mv = 16$ QSO.
Then, (\ref{eq:A11}) provides, respectively, the fundamental noise of 
$\sim (0.2-1.0)\times10^{-5}$\, 
in $\tilde{\sigma}_{\Delta\alpha/\alpha}$ for one measurement
of a \ion{Fe}{ii} pair of lines.

The fundamental photon noise limits for \ion{Fe}{ii} pairs from the
\zabs = 1.839 system are given in Table~4.  
The photon noise is computed with the
best-fitting parameters $N_1 = 1.335\times10^{13}$ \cm\,
and $b_1 = 5.25$ \kms.

\section{Discussion and Outlook}

\subsection{\daa\, from different studies}

Measurements from the present paper
can be combined with the previous \ion{Fe}{ii} sample from
the \zabs = 1.15 system toward \object{HE 0515--4414} 
(QRL) to increase statistics.
The normalized distribution ($\int\cdot\, d$\daa = 1)
of the resulting 35 \daa values is plotted in Fig.~6 (histogram) 
along with two other recently published results of MFWDPW and CSPA which are shown
by the dashed and dotted curves, respectively, assuming that the measured
\daa are normally distributed with the sample means and standard deviations
published in these papers. The vertical lines in this figure
mark the centers of the corresponding distributions.
The distribution shown by the histogram has the sample mean
$\langle$\daa$\rangle$ = $(0.4\pm1.5_{\rm stat})\times10^{-6}$ and the median
(\daa)$_{\rm med} = 0.1\times10^{-6}$. 

It is to be noted that at a given redshift  
the sample mean $\langle$\daa$\rangle$ should 
be the same within the uncertainty interval
independently on the method or the sample used -- provided
the data are free from any systematics. 
The results presented in Fig.~6 show, however, that
$\langle$\daa$\rangle_{\rm Keck/HIRES} \neq$
$\langle$\daa$\rangle_{\rm VLT/UVES}$.
The sample means of CSPA and our \ion{Fe}{ii} ensemble are in good agreement but
they differ from that of MFWDPW at the 95\% significance level according to 
the $t$-test\footnote{The application of the $t$-test to two normally
distributed and independent random variables with 
{\it different} variances is described in 
Bol'shev \& Smirnov (1983).}.  This discrepancy points to the systematic shift
which mimics the effect of varying $\alpha$ in the Keck/HIRES spectra.
To clarify the origin of this systematic shift, one needs more accurate measurements,
which can be carried out with higher spectral resolution
and with more homogeneous samples.

The comparison of the distribution widths in Fig.~6 reveals 
that the standard deviation in the CSPA sample is exceptionally 
small. For example, Fig.~1(b) in CSPA, where the 
accuracy of wavelength calibration is checked through the relative velocity
shifts, $\Delta v$, between the \ion{Fe}{ii} $\lambda2344$ and 
$\lambda2600$ lines\footnote{The mean $\langle \Delta v \rangle$ should be
consistent with zero in the case of good calibration since the sensitivity
coefficients ${\cal Q}$(\ion{Fe}{ii} $\lambda2344$)~$\simeq
{\cal Q}$(\ion{Fe}{ii} $\lambda2600$).},
shows the dispersion of
$\sigma_{\Delta v} \simeq 0.4$ \kms. 
This uncertainty in wavelength calibration transforms into the error
$\sigma_{\Delta\alpha/\alpha} \sim 2\times10^{-5}$ [see eq.(12) in L04], i.e.,
in order to reach the error of the mean 
$\sigma_{\langle \Delta \alpha/\alpha \rangle}
\sim 0.6\times10^{-6}$ (CSPA), one needs a sample of the size $n \sim 1100$,
which is not the case.
Thus, the error of the mean 
$\sigma_{\langle \Delta \alpha/\alpha \rangle}$
estimated by CSPA is in some disagreement with their Fig.~1(b).
The scatter of \daa in the Keck sample is
about 2 times the $\sigma_{\rm rms}$ value of our combined \ion{Fe}{ii} sample.

\begin{figure}[t]
\vspace{0.0cm}
\hspace{0.0cm}\psfig{figure=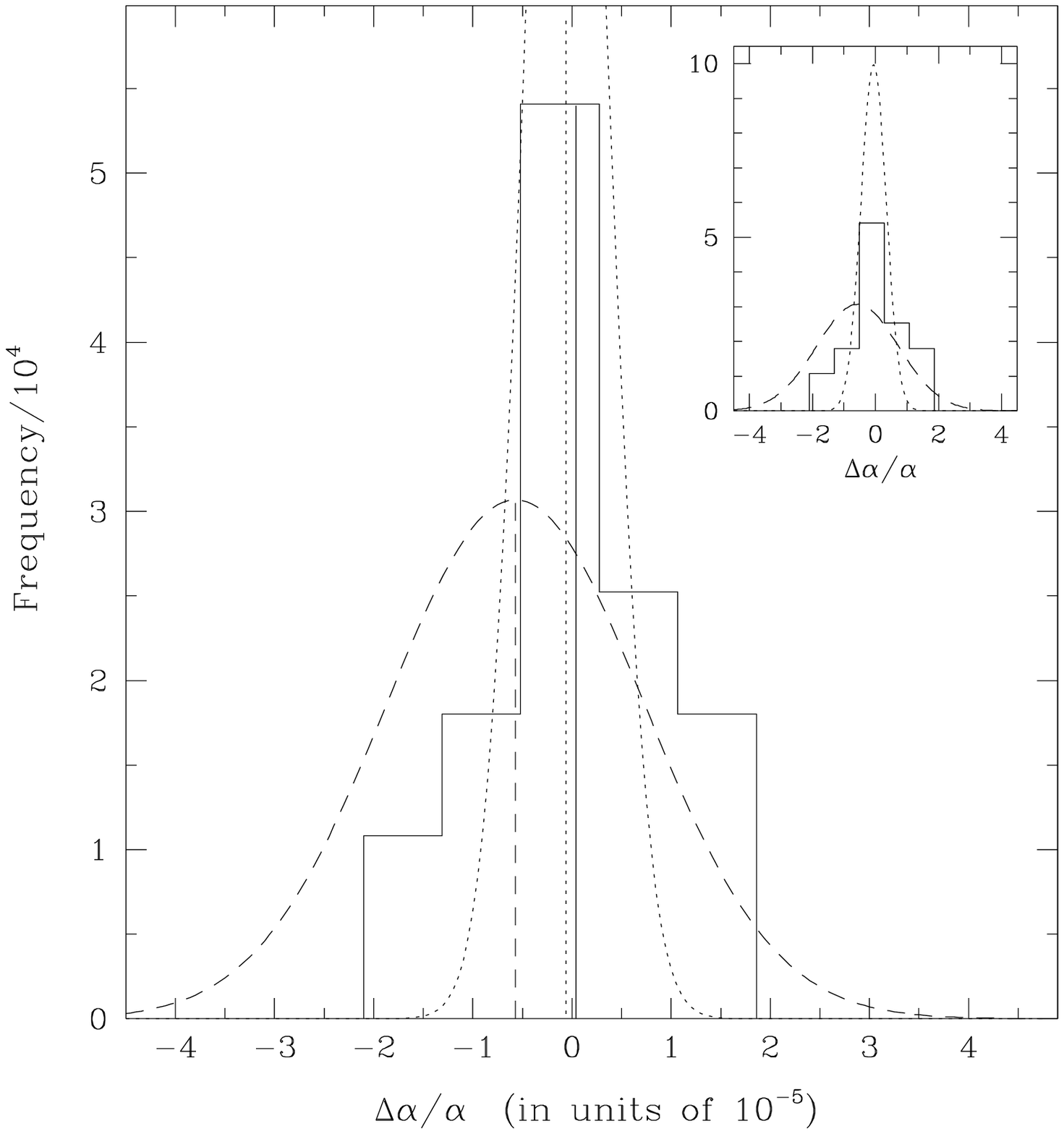,height=9.0cm,width=9.0cm}
\vspace{0.0cm}
\caption[]{The distribution of \daa (histogram)
from the \zabs = 1.839 and \zabs = 1.15 \ion{Fe}{ii}
systems toward \object{Q 1101--264} and \object{HE 0515--4414}, respectively. 
The distributions of \daa from MFWDPW (dashed curve) and CSPA (dotted curve) are
overplotted to demonstrate the discrepancy between the Keck/HIRES and VLT/UVES 
sample mean  $\langle$\daa$\rangle$ values.
Both MFWDPW and CSPA distributions are assumed to be normal with the means and
variances taken from the corresponding papers.
The vertical lines mark the centers of the distributions.
The revealed Keck--VLT discrepancy of $\langle$\daa$\rangle$  
is statistically significant at the 95\% confidence
level according to the $t$-test
}
\label{fig6}
\end{figure}

\subsection{Future prospects}

Future measurements of \daa from astronomical observations
should be carried out with high precision 
to check the present Oklo result that
$\alpha$ was larger in the past, \daa $\geq 4.5\times10^{-8}$ 
(Lamoreaux \& Torgerson 2004). 
To reach the Oklo scale of $\alpha$-variation, one needs the accuracy of at least
$\sigma_{\Delta\alpha/\alpha} \sim 10^{-7}$, which is 8 times higher than the
accuracy set by the errors in the laboratory wavelengths (Sect.~4.1).
At first glance the Oklo level is unachievable in astronomical observations. 
However, applying the SIDAM 
method to a few \ion{Fe}{ii} systems observed at different redshifts, one can 
omit the laboratory wavelengths from \daa\, calculations.
Such {\it self-calibrating} procedure, described by BSS,  
implies that the measured values of $R(t)$, eq.(\ref{eq:A7}),
can be fitted to a linear function of cosmic time
given in the form [cf. eq.(6) in BSS]:
\begin{equation}
R(t) = R(0)\,(1 + StH_0)\; ,
\label{eq:C1}
\end{equation}
where $R(0)$ is any convenient constant,
$H_0$ is the Hubble constant, and the slope $S$ is equal in our case to
\begin{equation}
S = \frac{\kappa}{H_0\,\alpha}\,\left(\frac{d\alpha}{dt}\right)\; .
\label{eq:C2}
\end{equation}
The constant $\kappa$ is defined through eq.(\ref{eq:A10}):
$\kappa = 4({\cal Q}_1 - {\cal Q}_2)(\eta + 1)/\eta(\eta + 2)$.
For the pair \ion{Fe}{ii} $\lambda\lambda1608, 2586$, $\kappa \simeq -1/4$.

Equation~(\ref{eq:B4}) shows that for narrow lines ($\xi < \sqrt{2}$)
the error
$\sigma_\lambda$ decreases almost quadratically with 
decreasing wavelength bin per pixel
if $N_{\rm e}$ is fixed.
Therefore, if an efficient spectrograph with ten times the UVES
dispersion and superior stability can be coupled to 
a 100~m class telescope, one would expect
$\sigma^{\rm lim}_\lambda \sim$ 0.03\,-\,0.05 m\AA, and, thus, 
the precision of $\sim 10^{-7}$ in the \daa measurements  
can be achieved.  Then, with good statistics the Oklo result can be checked
at different redshifts.

If $\alpha$ were indeed larger in the past, then the 
logarithmic derivative of $\alpha(t)$ in eq.(\ref{eq:C2})
is negative (since $t$ decreases with increasing redshift) and we
would observe a positive slope $S$. Otherwise, if $\alpha$ were smaller in the past, 
$S$ would be negative. 

\section{Conclusions}

The main results of the present paper are as follows:
\begin{enumerate}
\item A data reduction procedure is designed to control the
accuracy of the wavelength scale calibration and to 
bypass the influence of
the spectral shifts due to instabilities in the instrument
which perturb the narrow absorption-line profiles.
\item A single ion method is  proposed to minimize the influence of the
systematic shifts of the line centroids to the accuracy of the 
\daa\, measurements. 
\item The mean value 
$\langle$\daa$\rangle = (2.4\pm3.8_{\rm stat})\times10^{-6}$
is obtained from the analysis of
the \zabs = 1.839 \ion{Fe}{ii} system toward
\object{Q 1101--264}.
Combination of this measurement with the \zabs = 1.15 \ion{Fe}{ii} system
toward \object{HE 0515--4414} (QRL) 
increases the sample size to 35 \ion{Fe}{ii} pairs and
provides $\langle$\daa$\rangle = (0.4\pm1.5_{\rm stat})\times10^{-6}$. 
\item We confirm that the VLT/UVES estimations of \daa differ significantly from
the Keck/HIRES result, and that the systematic shift is present in the Keck data
with a probability of 95\%.
\item The fundamental noise limitation in the \daa measurement is defined and 
calculated for the \zabs = 1.839 \ion{Fe}{ii} system. It is shown
that a typical VLT/UVES fundamental noise limit is
$\sigma^{\rm lim}_{\Delta\alpha/\alpha} \sim 1\times10^{-5}$ for 
a pair of \ion{Fe}{ii} lines observed in a spectrum of a relatively bright QSO  
($V = 16.0$).
This estimation is robust since the inclusion of other metal transitions (like
\ion{Mg}{ii} $\lambda\lambda2796, 2803$; \ion{Al}{ii} $\lambda1670$;
\ion{Si}{ii} $\lambda\lambda1526, 1808$)
cannot improve significantly the error  
$\sigma^{\rm lim}_{\Delta\alpha/\alpha}$ due to lower sensitivity
to the variation of $\alpha$. 
From this point of view the dispersion
$\sigma_{\Delta\alpha/\alpha} = 0.4\times10^{-5}$ 
and the error of the mean 
$\sigma_{\langle \Delta\alpha/\alpha\rangle } = 0.6\times10^{-6}$
found in CSPA 
seems to be several times underestimated.
\item We suggest that with a spectrograph of $\sim 10$ times the UVES 
dispersion and superior stability
the accuracy high enough to probe the Oklo result,
$\Delta\alpha/\alpha \geq 4.5\times10^{-8}$, 
can be attained in future at new giant telescopes.
\end{enumerate}

\begin{acknowledgements}
The authors are indebted to Ralf Quast for his valuable comments.
S.A.L. gratefully acknowledges the hospitality of Osservatorio Astronomico
di Trieste where this work was performed under the program
COFIN 02 N 2002027319-001.
The work of S.A.L. is supported by
the RFBR grant No. 03-02-17522 and by the RLSS grant 1115.2003.2.
\end{acknowledgements}

\end{document}